\documentclass[11pt,openany]{article}
\usepackage{latexsym}
\usepackage{amsmath}
\usepackage{amssymb}
\usepackage{cite}
\usepackage{flafter}
\usepackage{epsfig}
\usepackage{graphicx}
\usepackage{bm}
\usepackage[mathscr]{eucal}
\input{epsf}
\usepackage{fancyhdr}
\renewcommand{\theequation}{\thesection.\arabic{equation}}
\begin{document}

\title{{\Large {\bf Quantum interference effect on the density of
states in disordered $d$-wave superconductors}}}
\author{Y.~H.~Yang,$^{1,2}$ Y.~G.~Wang,$^2$ M.~Liu,$^2$ and D.~Y.~Xing$^1$\\
\em\small $^{1}$National Laboratory of Solid State
Microstructures, Nanjing University, Nanjing 210008,
China\\\em\small $^{2}$Department of Physics, Southeast
University, Nanjing 210096, China}
\date{}
\maketitle
\begin{abstract}
{\large The quantum interference effect on the quasiparticle
density of states (DOS) is studied with the diagrammatic technique
in two-dimensional $d$-wave superconductors with dilute
nonmagnetic impurities both near the Born and near the unitary
limits. We derive in details the expressions of the Goldstone
modes (cooperon and diffuson) for quasiparticle diffusion. The DOS
for generic Fermi surfaces is shown to be subject to a quantum
interference correction of logarithmic suppression, but with
various renormalization factors for the Born and unitary limits.
Upon approaching the combined limit of unitarity and nested Fermi
surface, the DOS correction is found to become a $\delta$-function
of the energy, which can be used to account for the resonant peak
found by the numerical studies.}
\end{abstract}
\vspace{0.2in} PACS numbers: 74.25.Jb,  71.20.-b, 73.20.Fz

\begin{center}
{\section*{{\bf I. Introduction}}}
\end{center}
\setcounter{section}{1}
\setcounter{equation}{0}

Since the experiments revealed the $d$-wave symmetry of the order
parameter in cuprate superconductors~\cite{1}, the physics of
low-energy quasiparticle excitations in disordered two-dimensional
(2D) $d$-wave superconductors has been a subject of ongoing
intensive research~\cite{2}. The characteristic feature of the
$d_{x^2-y^2}$-wave pairing state is the existence of four nodal
points where the order parameter vanishes. In the vicinity of the
gap nodes there exist low-lying Dirac-type quasiparticle
excitations. An understanding of the disorder effect induced by
randomly-distributed impurities on these low-energy states is
essential for the elucidation of the thermodynamic and transport
properties of disordered $d$-wave superconductors. During the
years a number of theoretical approaches, such as the
self-consistent approximation schemes~[3--9], non-perturbative
methods~[10--17], and numerical studies~[18-22], have been
developed to calculate the quasiparticle density of states (DOS)
in the presence of disorder. Unfortunately, many of these theories
contradict each other. The DOS at zero energy was variously
predicted to be finite~[3--10], vanishing~[11--14,18,21], and
divergent~[15,20]. Recently, it has been made clear
that~[16,17,19,22] much of this controversy roots in the fact that
the $d$-wave superconductor is fundamentally sensitive to the
details of disorder, as well as to certain symmetries of the
normal state Hamiltonian.

In order to clarify the physics of the various asymptotic results
for the DOS, Yashenkin et al.~\cite{23} analyzed the diffusive
modes in disordered $d$-wave superconductors and calculated the
weak-localization correction to the DOS with the diagrammatic
technique. It is well known that the weak localization in electron
systems is a manifestation of the quantum interference (QI)
effect, which stems from the existence of the Goldstone modes
(cooperon and diffuson)~\cite{24}. As pointed out by Altland and
Zirnbauer~\cite{25}, the Goldstone modes in a superconductor have
different features from those in a normal metal, for the local
particle-hole symmetry of the superconducting state gives rise to
a combination of impurity- and Andreev-scattering processes. As a
result, every cooperon or diffuson mode in the retarded-advanced
(RA) channel entails a corresponding mode in the retarded-retarded
(RR) or advanced-advanced (AA) channel. In the combined limit of
the unitarity and nested Fermi surface (the UN limit), each of
these $0$-mode cooperon and diffuson has a $\pi$-mode counterpart
due to the global particle-hole symmetry~\cite{23,26}. For
disordered $d$-wave superconductors, it was found~\cite{23} that
in generic situations the quasiparticle DOS is subject to a
logarithmic weak-localization correction due to the 0-mode
cooperon, and the existence of diffusive $\pi$-modes can give rise
to a finite, or even divergent zero-energy DOS upon approaching
the UN limit.

The weak-localization calculations carried out by Yashenkin et al.
appear to capture the physical origin of the discrepancies between
predicted low-energy quasiparticle DOS. Furthermore, the
quasiparticle weak-localization effect was also suggested to have
important influences on transport properties such as the
electrical~\cite{6}, spin~\cite{13}, thermal~\cite{27}, and
optical~\cite{28} conductivities of $d$-wave superconductors.
While the QI effects have been widely investigated for disordered
normal metals~\cite{24}, a corresponding theory for random Dirac
fermions in superconducting cuprates is far from well established,
and thus highly deserves further development.

This paper presents an intensive study of the QI effect on the
quasiparticle DOS in 2D $d$-wave superconductors with dilute
nonmagnetic impurities both near the Born and near the unitary
limits. These two limiting cases are considered to be closely
related with the disorder effects in cuprate superconductors. It
is reasonable that the disorder due to defects off the
copper-oxygen plane may be treated in the Born approximation and
that defects in the plane may be in the unitary limit~\cite{6}.
Albeit sharing certain aspects with Ref.~\cite{23}, we further
develop the weak-localization theory in $d$-wave superconductors,
and obtain some new results in this paper. First, the expressions
of the Goldstone modes for quasiparticle diffusion are derived in
details both near the Born and near the unitary limits. Second, we
calculate the additional contributions to the DOS from those
lowest-order self-energy diagrams containing non-singular ladders,
which were not taking into account previously. These diagrams are
shown to give rise to various renormalization factors for the DOS
correction in the Born and unitary limits. Third, by taking into
account a new nontrivial self-energy diagram with the $\pi$-mode
diffuson, we show that the QI correction to the DOS becomes a
$\delta$-function of the energy upon approaching the UN limit.
This result can be used to account for the resonant peak found by
the previous numerical studies~\cite{19,20}.

The structure of this paper is as follows. In Sec.~II, the
commonly used self-consistent $T$-matrix approximation (SCTMA) is
described for a weakly-disordered $d_{x^2-y^2}$-wave
superconductor. Using the SCTMA, we derive in Sec.~III the
expressions of the 0-mode and $\pi$-mode cooperons and diffusons.
The QI correction to the quasiparticle DOS is calculated in
Sec.~IV, and the conclusions are summarized in Sec.~V. The
Appendix provides some mathematical formulas, which are useful for
our derivations.

\vspace{0.5cm}
\begin{center}
{\section*{{\bf II. SCTMA for d-wave superconductors}}}
\end{center}
\setcounter{section}{2} \setcounter{equation}{0} \vspace{-0.8cm}

Let us consider a most extensively studied model for a 2D
$d_{x^2-y^2}$-wave superconductor. In the tight-binding
approximation, the normal-state dispersion of a square lattice is
given by $\xi_{\bm k}=-t(\cos{k_xa}+\cos{k_ya})-\mu$ where $a$ is
the lattice constant, $t$ is the nearest-neighbor hopping
integral, and $\mu$ is the chemical potential. The nested Fermi
surface corresponds to the half-filling case ($\mu=0$). The order
parameter of the $d_{x^2-y^2}$-wave pairing state can be expressed
by $\Delta_{\bm k}=\Delta_0(\cos{k_xa}-\cos{k_ya})$. The four gap
nodes are given by ${\bm k}_n=\pm(k_0,\pm k_0)$ with
$k_0=(1/a)\arccos{(-\mu/2t)}$, which satisfy $\xi_{{\bm
k}_n}=\Delta_{{\bm k}_n}=0$. In the vicinity of these nodes, the
quasiparticle spectrum can be linearized as $\epsilon_{\bm
k}=\sqrt{\xi^2_{\bm k}+\Delta^2_{\bm k}}\approx\sqrt{({\bm
v}_f{\bm\cdot}\tilde{\bm k})^2+({\bm v}_g{\bm\cdot}\tilde{\bm
k})^2}$, where ${\bm v}_f=(\partial\xi_{\bm k}/\partial{\bm
k})_{{\bm k}_n}$, ${\bm v}_g=(\partial\Delta_{\bm k}/\partial{\bm
k})_{{\bm k}_n}$, and $\tilde{\bm k}$ is the momentum measured
from the node ${\bm k}_n$. A direct calculation yields
$v_f=v_gt/\Delta_0=\sqrt{2}ta\sqrt{1-(\mu/2t)^2}$.

In the presence of randomly-distributed nonmagnetic impurities,
the time-reversal and spin-rotational symmetries remain preserved.
Thus the system belongs to symmetry class CI in the classification
of Ref.~\cite{25}. The impurity potential is assumed to be
point-like, meaning that the intra- and inter-node scatterings are
described by a single potential strength $V$. For a low impurity
concentration $n_i$, the quasiparticle self-energy in the
SCTMA~\cite{23} can be expressed in the Nambu spinor
representation as
\begin{equation}
\Sigma^{R(A)}(\epsilon)=n_iT^{R(A)}(\epsilon)=(\lambda\epsilon\mp
i\gamma)\tau_0+\eta\gamma\tau_3,
\end{equation}
for $|\epsilon|\ll\gamma$. Here $\lambda$ is the mass
renormalization factor, $\gamma$ is the impurity-induced
relaxation rate, $\eta$ is a dimensionless parameter, $\eta\gamma$
represents the decrease of the chemical potential induced by the
impurity scatterings, and $\tau_0$ and $\tau_i$ $(i=1,2,3)$ stand
for the $2\times 2$ unity and Pauli matrices, respectively. A use
of Dyson's equation immediately yields the impurity-averaged
one-particle Green's functions as
\begin{equation}
G^{R(A)}_{\bm k}(\epsilon)=\frac{[(1-\lambda)\epsilon\pm
i\gamma]\tau_0+\Delta_{\bm k}\tau_1+\xi_{\bm
k}\tau_3}{[(1-\lambda)\epsilon\pm i\gamma]^2-\epsilon^2_{\bm k}},
\end{equation}
where the shift of chemical potential has been absorbed in $\mu$.
The zero-energy density of states is calculated as $
\rho_0=-(1/\pi)\mbox{Im}\sum_{\bm k}\mbox{Tr}G^R_{\bm
k}=4l\gamma/\pi^2v_fv_g$, where $G^{R(A)}_{\bm k}=G^{R(A)}_{\bm
k}(0)$ and $l=\ln(\Gamma/\gamma)>1$ with
$\Gamma\sim\sqrt{v_fv_g}/a$.

The parameters $\gamma$, $\lambda$, and $\eta$ can be evaluated
consistently via the $T$-matrix equation
\begin{equation}
T^{R(A)}(\epsilon)=V\tau_3+V\tau_3g^{R(A)}(\epsilon)T^{R(A)}(\epsilon),
\end{equation}
with $g^{R(A)}(\epsilon)=\sum_{\bm k}G_{\bm k}^{R(A)}(\epsilon)$.
Using Eq.~(2.2), we can show that
\begin{equation}
g^{R(A)}(\epsilon)=\frac{\pi\rho_0}{2\gamma}\left[(\lambda-1)(1-l^{-1})\epsilon\mp
i\gamma\right]\tau_0 +\left(V^{-1}-U^{-1}\right)\tau_3,
\end{equation}
for $|\epsilon|\ll\gamma$, where $U$ is the effective impurity
potential given by $U^{-1}=V^{-1}+\sum_{\bm k}\xi_{\bm
k}/(\epsilon^2_{\bm k}+\gamma^2)$. A substitution of Eqs.~(2.1)
and (2.4) into Eq.~(2.3) leads to
$\gamma=2n_i/\pi\rho_0(1+\eta^2)$,
$\lambda=(1-\eta^2)(l-1)/(\eta^2+2l-1)$, and $\eta=2/\pi\rho_0U$.

The Born limit corresponds to $\eta^2\gg 2l$, yielding
\begin{equation}\gamma=\frac{\pi}{2}n_i\rho_0U^2,\
\lambda=1-l;
\end{equation}
and the unitary limit corresponds to $\eta\to 0$, meaning that
\begin{equation}\gamma=\frac{2n_i}{\pi\rho_0},
\ \lambda=\frac{l-1}{2l-1}.
\end{equation}
Throughout this paper, we mainly consider the cases near either
the Born or the unitary limit. It is worthy to point out that the
values of the impurity potential $V$ driving the system into these
two limits are dependent on the band structure.

\vspace{0.5cm}
\begin{center}
{\section*{{\bf III. The diffusive modes}}}
\end{center}
\setcounter{section}{3} \setcounter{equation}{0}

Since the QI effect is related to the diffusive modes, we first
derive the expressions of the 0-mode and $\pi$-mode cooperons and
diffusons for a disordered $d$-wave superconductor.

\vspace{0.5cm}
\begin{center} {\subsection*{\bf A. $0$-mode
cooperon and diffuson}}
\end{center}
\vspace{-0.8cm}

The $0$-mode cooperon and diffuson exist both in RA and in RR
channels due to the local particle-hole symmetry $\tau_2G^R_{\bm
k}(\epsilon)\tau_2=-G^A_{\bm k}(-\epsilon)$. The ladder diagrams
for the cooperon are given by Fig. 1(b) in Ref.~\cite{23}. The
equation for $0$-mode cooperon can be expressed as
\begin{equation}
{\cal C}({\bm q};\epsilon,\epsilon')={\cal
W}(\epsilon,\epsilon')+{\cal W}(\epsilon,\epsilon'){\cal H}({\bm
q};\epsilon,\epsilon'){\cal C}({\bm q};\epsilon,\epsilon'),
\end{equation}
where the two-particle irreducible vertex ${\cal
W}(\epsilon,\epsilon')$ and the integral kernel ${\cal H}({\bm
q};\epsilon,\epsilon')$ are defined in the RR and RA channels as
\begin{equation}
{\cal W}(\epsilon,\epsilon')^{RR(A)}=n_iT^R(\epsilon)\otimes
T^{R(A)}(\epsilon')
\end{equation}
and
\begin{equation}
{\cal H}({\bm q};\epsilon,\epsilon')^{RR(A)}=\sum_{\bm k}G^R_{{\bm
q}-{\bm k}}(\epsilon)\otimes G^{R(A)}_{\bm k}(\epsilon').
\end{equation}
Equation (3.1) can be also expressed as
\begin{equation}
{\cal A}({\bm q};\epsilon,\epsilon'){\cal C}({\bm
q};\epsilon,\epsilon')={\cal W}(\epsilon,\epsilon'),
\end{equation}
where
\begin{equation}
{\cal A}({\bm q};\epsilon,\epsilon')={\cal I}-{\cal
W}(\epsilon,\epsilon'){\cal H}({\bm q};\epsilon,\epsilon'),
\end{equation}
with ${\cal I}=\tau_0\otimes\tau_0$. From Eqs.~(3.1)--(3.5), it
follows that one can make a decomposition of ${\cal
X}=\sum_{ij}X_{ij}\tau_i\otimes\tau_j$ for ${\cal X}={\cal
W},{\cal H},{\cal A}$, and ${\cal C}$. Substituting Eq.~(2.1) into
Eq. (3.2), we can easily obtain all the nonvanishing components of
${\cal W}(\epsilon,\epsilon')$ as
\begin{equation}
W(\epsilon,\epsilon')^{RR(A)}_{00}=\mp
\frac{2\gamma}{\pi\rho_0(1+\eta^2)}\left[1+i\frac{\lambda}{\gamma}(\epsilon\pm\epsilon')\right],
\end{equation}
\begin{equation}
W(\epsilon,\epsilon')^{RR(A)}_{33}=
\frac{2\gamma\eta^2}{\pi\rho_0(1+\eta^2)},
\end{equation}
\begin{equation}
W(\epsilon,\epsilon')^{RR(A)}_{03}=-i
\frac{2\gamma\eta}{\pi\rho_0(1+\eta^2)}\left(1+i\frac{\lambda}{\gamma}\epsilon\right),
\end{equation}
and
\begin{equation}
W(\epsilon,\epsilon')^{RR(A)}_{30}=\mp i
\frac{2\gamma\eta}{\pi\rho_0(1+\eta^2)}\left(1\pm
i\frac{\lambda}{\gamma}\epsilon'\right).
\end{equation}
It then follows that the dominant component of ${\cal
W}(\epsilon,\epsilon')$ near the Born limit ($\eta^2\gg2l$) is
$W(\epsilon,\epsilon')_{33}$, while that near the unitary limit
($\eta^2\ll1$) is $W(\epsilon,\epsilon')_{00}$.

Now let us calculate ${\cal H}({\bm q};\epsilon,\epsilon')$ and
${\cal A}({\bm q};\epsilon,\epsilon')$. For small values of $\bm
q$, $\epsilon$, and $\epsilon'$, we have
\begin{equation}
G^R_{{\bm q}-{\bm k}}(\epsilon)\approx G^R_{\bm
k}+\epsilon\frac{\partial}{\partial \epsilon'}G^R_{\bm
k}(\epsilon')\mid_{\epsilon'=0}-{\bm q}\cdot\nabla G^R_{\bm
k}+\frac{1}{2}{\bm q}{\bm q}:\nabla\nabla G^R_{\bm k}
\end{equation}
and
\begin{equation}
G^{R(A)}_{\bm k}(\epsilon')\approx G^{R(A)}_{\bm
k}+\epsilon'\frac{\partial}{\partial \epsilon}G^{R(A)}_{\bm
k}(\epsilon)\mid_{\epsilon=0}.
\end{equation}
Substituting Eqs. (3.10) and (3.11) into Eq. (3.3), and using Eqs.
(A1)--(A4) in the Appendix, we can show that all the nonvanishing
components of ${\cal H}({\bm q};\epsilon,\epsilon')$ are given by
\begin{equation}
H({\bm
q};\epsilon,\epsilon')^{RR(A)}_{00}=\mp\frac{\pi\rho_0}{4l\gamma}\left(
1-\frac{v^2_f+v^2_g}{12\gamma^2}q^2\right),
\end{equation}
\begin{equation}
H({\bm
q};\epsilon,\epsilon')^{RR(A)}_{11}=\frac{\pi\rho_0}{8l\gamma}\left[(2l-1)
+i\frac{1-\lambda}{\gamma}(\epsilon\pm\epsilon')-\frac{v^2_f+3v^2_g}{12\gamma^2}q^2\right],
\end{equation}
and
\begin{equation}
H({\bm
q};\epsilon,\epsilon')^{RR(A)}_{33}=\frac{\pi\rho_0}{8l\gamma}\left[(2l-1)
+i\frac{1-\lambda}{\gamma}(\epsilon\pm\epsilon')-\frac{v^2_g+3v^2_f}{12\gamma^2}q^2\right].
\end{equation}
The diagonal components of ${\cal A}({\bm q};\epsilon,\epsilon')$
can be easily calculated by a substitution of Eqs. (3.6), (3.7),
and (3.12)--(3.14) into Eq. (3.5), yielding
\begin{eqnarray}
A({\bm
q};\epsilon,\epsilon')^{RR(A)}_{00}=-\frac{1}{4l(1+\eta^2)}\Big[-2(2l-1)-(2l+1)\eta^2\nonumber\\
+i\frac{2\lambda+(1-\lambda)\eta^2}{\gamma}(\epsilon\pm\epsilon')
-\frac{2v^2_f+2v^2_g+(3v_f^2+v_g^2)\eta^2}{12\gamma^2}q^2\Big],
\end{eqnarray}
\begin{eqnarray}
A({\bm
q};\epsilon,\epsilon')^{RR(A)}_{11}=\pm\frac{1}{4l(1+\eta^2)}\Big[(2l-1)
+i\frac{2\lambda(l-1)+1}{\gamma}(\epsilon\pm\epsilon')-\frac{v^2_f+3v^2_g}{12\gamma^2}q^2\Big],
\end{eqnarray}
\begin{equation}
A({\bm
q};\epsilon,\epsilon')^{RR(A)}_{22}=\frac{\eta^2}{4l(1+\eta^2)}\Big[(2l-1)
+i\frac{1-\lambda}{\gamma}(\epsilon\pm\epsilon')-\frac{v^2_f+3v^2_g}{12\gamma^2}q^2\Big],
\end{equation}
and
\begin{eqnarray}
A({\bm
q};\epsilon,\epsilon')^{RR(A)}_{33}=\pm\frac{1}{4l(1+\eta^2)}\Big[(2l-1)+2\eta^2\nonumber\\
+i\frac{2\lambda(l-1)+1}{\gamma}(\epsilon\pm\epsilon')
-\frac{v^2_g+3v^2_f+(2v_f^2+2v_g^2)\eta^2}{12\gamma^2}q^2\Big].
\end{eqnarray}

We note that all the non-diagonal components of ${\cal H}({\bm
q};\epsilon,\epsilon')$ are vanishing, and those of ${\cal
W}(\epsilon,\epsilon')$ are negligible near the Born or unitary
limit. Then Eq.~(3.1) or (3.4) indicates that only the diagonal
components of ${\cal C}({\bm q};\epsilon,\epsilon')$ may be
singular for these two limits. As a result, the 0-mode cooperon
can be expressed as ${\cal C}({\bm
q};\epsilon,\epsilon')=\sum_iC({\bm
q};\epsilon,\epsilon')_{ii}\tau_i\otimes \tau_i$. As will be shown
below, all $C({\bm q};\epsilon,\epsilon')_{ii}$ are of diffusive
poles. Obviously, only the diagonal components of ${\cal A}({\bm
q};\epsilon,\epsilon')$ and ${\cal W}(\epsilon,\epsilon')$ are
needed for the calculation of 0-mode cooperon, and thus Eq.~(3.4)
becomes equivalent to the following group of equations:
\begin{equation}
A_{00}C_{00}+A_{11}C_{11}+A_{22}C_{22}+A_{33}C_{33}=W_{00},
\end{equation}
\begin{equation}
A_{00}C_{11}+A_{11}C_{00}-A_{22}C_{33}-A_{33}C_{22}=0,
\end{equation}
\begin{equation}
A_{00}C_{22}-A_{11}C_{33}+A_{22}C_{00}-A_{33}C_{11}=0,
\end{equation}
\begin{equation}
A_{00}C_{33}-A_{11}C_{22}-A_{22}C_{11}+A_{33}C_{00}=W_{33},
\end{equation}
where the arguments (${\bm q}$, $\epsilon$, and $\epsilon'$) of
$A_{ii}$, $C_{ii}$, and $W_{ii}$ have been omitted.

Near the Born limit ($\eta^2\gg 2l$), the singular terms in Eqs.
(3.19)--(3.22) satisfy the following relations:
\begin{equation}
(2l+1)C_{00}^{RR(A)}+(2l-1)C_{22}^{RR(A)}\pm2C_{33}^{RR(A)}=0,
\end{equation}
\begin{equation}
(2l+1)C_{11}^{RR(A)}-(2l-1)C_{33}^{RR(A)}\mp2C_{22}^{RR(A)}=0,
\end{equation}
\begin{equation}
(2l+1)C_{22}^{RR(A)}+(2l-1)C_{00}^{RR(A)}\mp2C_{11}^{RR(A)}=0,
\end{equation}
\begin{equation}
(2l+1)C_{33}^{RR(A)}-(2l-1)C_{11}^{RR(A)}\pm2C_{00}^{RR(A)}=0,
\end{equation}
the only solution of which is given by
\begin{equation}
C_{00}^{RR(A)}=\mp C_{11}^{RR(A)}=-C_{22}^{RR(A)}=\mp
C_{33}^{RR(A)}.
\end{equation}
Substituting Eqs. (3.7), (3.15)--(3.18), and (3.27) into Eq.
(3.22), and using Eq.~(2.5), we can show that the terms of order
$\eta^{-2}$ cancel exactly out, leading to
\begin{equation}
C({\bm q};\epsilon,\epsilon')
^{RR(A)}_{00}=\mp\frac{4\gamma^2}{\pi\rho_0}\frac{1}{Dq^2-i(\epsilon\pm\epsilon')},
\end{equation}
where $D=(v_f^2+v_g^2)/4l\gamma$ is the quasiparticle diffusion
coefficient. Combining Eq.~(3.27) with Eq.~(3.28), and noting that
the $0$-mode diffuson has the same expression as that of the
$0$-mode cooperon due to the time-reversal symmetry, we obtain
\begin{equation}
{\cal C}({\bm q};\epsilon,\epsilon')^{RR(A)}={\cal D}({\bm
q};\epsilon,\epsilon')^{RR(A)}
=\frac{4\gamma^2}{\pi\rho_0}\frac{1}{Dq^2-i(\epsilon\pm\epsilon')}
\Big(\mp\tau_0\otimes\tau_0+\tau_1\otimes\tau_1
\pm\tau_2\otimes\tau_2+\tau_3\otimes\tau_3\Big),
\end{equation}
which is in agreement with that appearing in Ref.~\cite{26} (aside
from a disputed pre-factor). Similarly, we can show that
Eq.~(3.29) is also valid near the unitary limit ($\eta^2\ll 1$).
The above evaluations indicate that any small deviations from
either limit do not make the Goldstone 0-modes gapped. However, in
the intermediate region far from these two limits, the
non-diagonal components of ${\cal W}(\epsilon,\epsilon')$ cannot
be neglected, and thus ${\cal C}({\bm q};\epsilon,\epsilon')$ may
contain some singular non-diagonal components.

\begin{center}
{\subsection*{\bf B. $\pi$-mode cooperon and diffuson}}
\end{center}

In the UN limit, there exist the additional $\pi$-mode cooperon
and diffuson due to the global particle-hole symmetry~\cite{23,26}
$\tau_2G^{R(A)}_{\bm k}(\epsilon)\tau_2=G^{R(A)}_{{\bm Q}+{\bm
k}}(\epsilon)$, with ${\bm Q}=\pm(\pi/a,\pm\pi/a)$ the nesting
vector. Any small deviations from this combined limit can be shown
to make the Goldstone $\pi$-modes gapped. The ladder diagrams for
the $\pi$-mode cooperon can be obtained from those of the $0$-mode
cooperon by replacing ${\bm q}$ by ${\bm Q}+{\bm q}$. The equation
for the $\pi$-mode cooperon is given by
\begin{equation}
{\cal A}_\pi({\bm q};\epsilon,\epsilon'){\cal C}_\pi({\bm
q};\epsilon,\epsilon')={\cal W}(\epsilon,\epsilon'),
\end{equation}
where
\begin{equation}
{\cal A}_\pi({\bm q};\epsilon,\epsilon')={\cal I}-{\cal
W}(\epsilon,\epsilon'){\cal H}_\pi({\bm q};\epsilon,\epsilon'),
\end{equation}
with
\begin{equation}
{\cal H}_\pi({\bm q};\epsilon,\epsilon')^{RR(A)}=\sum_{\bm k}G
^R_{{\bm Q}+{\bm q}-{\bm k}}(\epsilon)\otimes G^{R(A)}_{\bm
k}(\epsilon').
\end{equation}

In order to calculate ${\cal H}_\pi({\bm q};\epsilon,\epsilon')$
and ${\cal A}_\pi({\bm q};\epsilon,\epsilon')$, one needs to
exploit the relations $\Delta_{{\bm Q}+{\bm k}}=-\Delta_{\bm k}$
and $\xi_{{\bm Q}+{\bm k}}=-\xi_{\bm k}-2\mu$. For a nearly-nested
Fermi surface $(|\mu|\ll \gamma)$, we have
\begin{eqnarray}
G ^R_{{\bm Q}+{\bm q}-{\bm k}}(\epsilon)\approx\tau_2G^R_{{\bm
q}-{\bm k}}(\epsilon)\tau_2+2\mu\frac{2\xi_{\bm
k}(i\gamma\tau_0-\Delta_{\bm k}\tau_1)+(\gamma^2+\Delta_{\bm
k}^2-\xi_{\bm k}^2)\tau_3}{(\gamma^2+\epsilon_{\bm
k}^2)^2}\nonumber\\+(2\mu)^2\frac{(\gamma^2+\Delta_{\bm
k}^2-3\xi_{\bm k}^2)(i\gamma\tau_0-\Delta_{\bm k}\tau_1)-\xi_{\bm
k}(3\gamma^2+3\Delta_{\bm k}^2-\xi_{\bm
k}^2)\tau_3}{(\gamma^2+\epsilon_{\bm k}^2)^3}.
\end{eqnarray}
Substituting Eqs. (3.11) and (3.33) into Eq. (3.32), and using
Eqs. (A1)--(A4), we can show that all the nonvanishing components
of ${\cal H}_\pi({\bm q};\epsilon,\epsilon')$ are given by
\begin{equation}
H_\pi({\bm q};\epsilon,\epsilon')^{RR(A)}_{00}=H({\bm
q};\epsilon,\epsilon')^{RR(A)}_{00}\pm\frac{\pi\rho_0\mu^2}{6l\gamma^3},
\end{equation}
\begin{equation}
H_\pi({\bm q};\epsilon,\epsilon')^{RR(A)}_{11}=-H({\bm
q};\epsilon,\epsilon')^{RR(A)}_{11}+\frac{\pi\rho_0\mu^2}{12l\gamma^3},
\end{equation}
\begin{equation}
H_\pi({\bm q};\epsilon,\epsilon')^{RR(A)}_{33}=-H({\bm
q};\epsilon,\epsilon')^{RR(A)}_{33}+\frac{\pi\rho_0\mu^2}{4l\gamma^3},
\end{equation}
and
\begin{equation}
H_\pi({\bm q};\epsilon,\epsilon')^{RR(A)}_{03}=\pm H_\pi({\bm
q};\epsilon,\epsilon')^{RR(A)}_{30}=-i\frac{\pi\rho_0\mu}{4l\gamma^2}.
\end{equation}
A substitution of Eqs.~(3.6)--(3.9) and (3.34)--(3.37) into
Eq.~(3.31) yields the diagonal components of ${\cal A}_\pi({\bm
q};\epsilon,\epsilon')$ as
\begin{eqnarray}
A_\pi({\bm
q};\epsilon,\epsilon')^{RR(A)}_{00}=-\frac{1}{4l(1+\eta^2)}\Big[-2(2l-1)-(6l-1)\eta^2
+i\frac{2\lambda-(1-\lambda)\eta^2}{\gamma}(\epsilon\pm\epsilon')\nonumber\\
-\frac{2v^2_f+2v^2_g-(3v_f^2+v_g^2)\eta^2}{12\gamma^2}q^2-\frac{(4-6\eta^2)\mu^2}{3\gamma^2}
-\frac{4\eta\mu}{\gamma}\Big],
\end{eqnarray}
\begin{eqnarray}
A_\pi({\bm
q};\epsilon,\epsilon')^{RR(A)}_{11}=\mp\frac{1}{4l(1+\eta^2)}\Big[(2l-1)
+i\frac{2\lambda(l-1)+1}{\gamma}(\epsilon\pm\epsilon')-\frac{v^2_f+3v^2_g}{12\gamma^2}q^2
-\frac{2\mu^2}{3\gamma^2}\Big],
\end{eqnarray}
\begin{equation}
A_\pi({\bm
q};\epsilon,\epsilon')^{RR(A)}_{22}=-\frac{\eta^2}{4l(1+\eta^2)}\Big[(2l-1)
+i\frac{1-\lambda}{\gamma}(\epsilon\pm\epsilon')-\frac{v^2_f+3v^2_g}{12\gamma^2}q^2
-\frac{2\mu^2}{3\gamma^2}\Big],
\end{equation}
and
\begin{eqnarray}
A_\pi({\bm
q};\epsilon,\epsilon')^{RR(A)}_{33}=\mp\frac{1}{4l(1+\eta^2)}\Big[(2l-1)-2\eta^2
+i\frac{2\lambda(l-1)+1}{\gamma}(\epsilon\pm\epsilon')\nonumber\\
-\frac{v^2_g+3v^2_f-(2v_f^2+2v_g^2)\eta^2}{12\gamma^2}q^2-\frac{(6-4\eta^2)\mu^2}{3\gamma^2}
-\frac{4\eta\mu}{\gamma}\Big].
\end{eqnarray}

Like the $0$-mode cooperon, the $\pi$-mode cooperon near the Born
or unitary limit can be also expressed as ${\cal C}_\pi({\bm
q};\epsilon,\epsilon')=\sum_iC_\pi({\bm
q};\epsilon,\epsilon')_{ii}\tau_i\otimes \tau_i$, with all
$C_\pi({\bm q};\epsilon,\epsilon')_{ii}$ assumed to be of
diffusive poles. Therefore, Eqs. (3.19)--(3.22) are also suitable
for the $\pi$-mode cooperon. Near the unitary limit ($\eta^2\ll
1$), the singular terms in Eqs. (3.19)--(3.22) (for the $\pi$-mode
cooperon) satisfy the following relations:
\begin{equation}
2C_{\pi00}^{RR(A)}\mp C_{\pi11}^{RR(A)}\mp C_{\pi33}^{RR(A)}=0,
\end{equation}
\begin{equation}
2C_{\pi11}^{RR(A)}\mp C_{\pi00}^{RR(A)}\pm C_{\pi22}^{RR(A)}=0,
\end{equation}
\begin{equation}
2C_{\pi22}^{RR(A)}\pm C_{\pi33}^{RR(A)}\pm C_{\pi11}^{RR(A)}=0,
\end{equation}
\begin{equation}
2C_{\pi33}^{RR(A)}\pm C_{\pi22}^{RR(A)}\mp C_{\pi00}^{RR(A)}=0,
\end{equation}
the solution of which is shown to be
\begin{equation}
C_{\pi00}^{RR(A)}=\pm C_{\pi11}^{RR(A)}=-C_{\pi22}^{RR(A)}=\pm
C_{\pi33}^{RR(A)}.
\end{equation}
Substituting Eqs. (3.6), (3.38)--(3.41) and (3.46) into Eq.
(3.19), and using Eq.~(2.6), we can show that
\begin{equation}
C_\pi({\bm
q};\epsilon,\epsilon')_{00}^{RR(A)}=\mp\frac{4\gamma^2}{\pi\rho_0}
\frac{1}{Dq^2-i(\epsilon\pm\epsilon')+2\delta},
\end{equation}
with
$\delta=2\eta^2\gamma+2\eta\mu/\gamma+\mu^2/l\gamma\ll\gamma$. The
present expression of $\delta$ supports the result estimated by
Yashenkin et al.~\cite{23}. Combining Eq.~(3.46) with Eq.~(3.47),
and noting that the $\pi$-mode diffuson has also the same
expression as that of the $\pi$-mode cooperon due to the
time-reversal symmetry, we obtain
\begin{eqnarray}
{\cal C}_\pi({\bm q};\epsilon,\epsilon')^{RR(A)}={\cal D}_\pi({\bm
q};\epsilon,\epsilon')^{RR(A)}\nonumber\\
=-\frac{4\gamma^2}{\pi\rho_0}\frac{1}{Dq^2-i(\epsilon\pm\epsilon')+2\delta}
\Big(\pm\tau_0\otimes\tau_0+\tau_1\otimes\tau_1
\mp\tau_2\otimes\tau_2+\tau_3\otimes\tau_3\Big).
\end{eqnarray}
Near the Born limit ($\eta^2\gg 2l$), one can easily show that
$C_{\pi ii}^{RR(A)}=0,~(i=0,1,2,3)$, indicating that the diffusive
$\pi$-modes exist only near the UN limit. Contrary to the
diffusive 0-modes, the Goldstone $\pi$-modes are gapped by any
small deviations from the UN limit measured by $\delta$. For the
situations far from the UN limit, the contributions of diffusive
$\pi$-modes to the QI effect are completely suppressed due to the
large gap.

\vspace{0.5cm}
\begin{center}
{\section*{{\bf IV. QI correction to the quasiparticle DOS}}}
\end{center}
\setcounter{section}{4} \setcounter{equation}{0} \vspace{-0.5cm}

The QI correction to the quasiparticle DOS can be calculated via
\begin{equation}
\Delta\rho(\epsilon)=-\frac{1}{\pi}{\rm Im}\sum_{\bm k}{\rm
Tr}\Delta G^R_{\bm k}(\epsilon),
\end{equation}
where $\Delta G^R_{\bm k}(\epsilon)$ represents the lowest-order
correction to the one-particle Green's function due to the
diffusive modes.

\begin{center}
{\subsection*{\bf A. Contribution of the 0-mode cooperon}}
\end{center}

The lowest-order self-energy diagrams with 0-mode cooperon are
depicted in Fig.~1.
\begin{figure}[htbp]
  \begin{center}
  \psfig{file=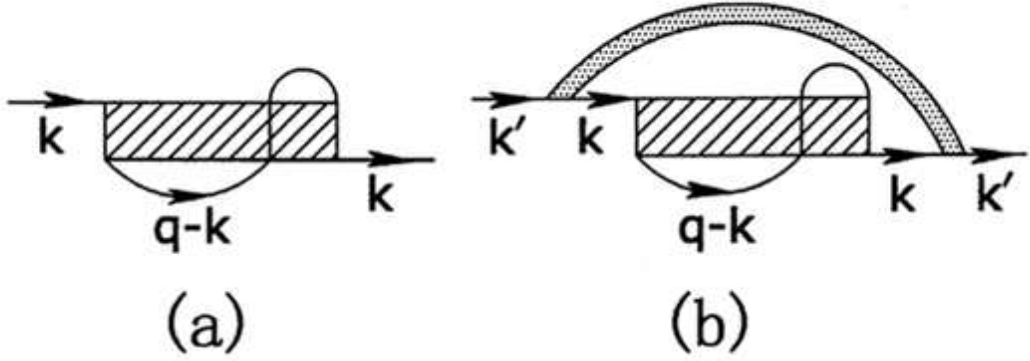,width=14cm,bbllx=41pt,bblly=360pt,bburx=572pt,bbury=563pt}
    \parbox{12.5cm}{\caption{\footnotesize The lowest-order self-energy diagrams with 0-mode cooperon
(shaded blocks). The grey block in Fig.~1(b) represents the
non-singular ladders. The corresponding diagrams with $\pi$-mode
cooperon can be obtained by the replacement of ${\bm
q}\rightarrow{\bm Q}+{\bm q}$ in Figs.~1(a) and 1(b).
   }}
   \end{center}
\end{figure}
The contribution of Fig.~1(a) has been calculated in
Ref.~\cite{23}. Figure 1(b) contains the non-singular ladders, its
contribution to the DOS can be shown to be the same order of
magnitude as that of Fig.~1(a). According to Eq.~(4.1), the
contribution of Fig.~1(a) to the DOS reads
\[
\rho(\epsilon)_{a}=-\frac{1}{\pi}{\rm Im}\sum_{{\bm k}{\bm
q}}\sum_iC({\bm q};\epsilon,\epsilon)^{RR}_{ii}{\rm
Tr}\left(\tau_iG^R_{{\bm q}-{\bm k}}\tau_iG^R_{\bm k}G^R_{\bm
k}\right).
\]
By means of Eq.~(A5), it can be rewritten as
\[
\rho(\epsilon)_{a}=-\frac{2}{\pi}{\rm Im}\sum_{\bm
q}\sum_i\left[{\cal C}({\bm q};\epsilon,\epsilon)^{RR}{\cal
M}_a\right]_{ii},
\]
with ${\cal M}_a=\sum_{\bm k}G^R_{\bm k}\otimes\left(G^R_{\bm
k}G^R_{\bm k}\right)$. The contribution of Fig.~1(b) can be
similarly evaluated, and the total contribution of Figs.~1(a) and
1(b) can be expressed by
\begin{equation}
\rho(\epsilon)_{\rm coop}=-\frac{2}{\pi}{\rm Im}\sum_{\bm
q}\sum_i\left[{\cal C}({\bm q};\epsilon,\epsilon)^{RR}{\cal
M}\right]_{ii},
\end{equation}
with
\begin{equation}
{\cal M}=\sum_{\bm k}G^R_{\bm k}\otimes(G^R_{\bm k}KG^R_{\bm k}).
\end{equation}
Here the matrix $K$ satisfies
\begin{equation}
K=\tau_0+n_i\sum_{\bm k}T^RG^R_{\bm k}KG^R_{\bm k}T^R,
\end{equation}
with $T^{R(A)}=T^{R(A)}(0)$. Substituting Eqs.~(2.1) and (2.2)
into Eq.~(4.4), and using Eqs.~(A1)--(A4), we can show that
\begin{equation}
K=(l/\zeta)\tau_0,
\end{equation}
where
\begin{eqnarray}
\zeta=\left\{\begin{array}{cc}1,~&{\rm for ~the ~Born ~limit},\\
     2l-1,~&{\rm
     for ~the ~unitary ~limit}.\end{array}\right.
\end{eqnarray}
Equation (4.5) leads to ${\cal M}=(l/\zeta){\cal M}_a$, and hence
$\rho(\epsilon)_{\rm coop}=(l/\zeta)\rho(\epsilon)_a$, indicating
that the contribution of Fig.~1(b) just renormalizes the
pre-factor of $\rho(\epsilon)_a$. A substitution of Eqs.~(2.2) and
(4.5) into Eq.~(4.3) yields
\begin{equation}
{\cal M}=-i\frac{l}{2\pi\zeta
v_fv_g\gamma}\Big(\tau_1\otimes\tau_1+\tau_3\otimes\tau_3\Big).
\end{equation}
Substituting Eqs.~(3.29) and (4.7) into Eq.~(4.2), and noting that
the upper cutoff of $q$ can be set to be $1/l_e$ with
$l_e=\sqrt{D/2\gamma}$ the mean-free path, we obtain
\begin{eqnarray}
\frac{\rho(\epsilon)_{\rm coop}}{\rho_0}&=&-\frac{2\pi}{\alpha
\zeta}{\rm Re}\sum_{\bm
q}\frac{D}{Dq^2-i2\epsilon}\nonumber\\&=&-\frac{1}{2\alpha
\zeta}\ln\frac{\gamma}{|\epsilon|}.
\end{eqnarray}
where $\alpha=(v_f^2+v_g^2)/2v_fv_g$.

\vspace{0.5cm}
\begin{center}
{\subsection*{\bf B. Vanishing contribution of the 0-mode
diffuson}}
\end{center}
\vspace{0.cm}

The lowest-order self-energy diagrams with the $0$-mode diffuson
are given by Figs.~2(a) and 2(b), and the vertex correction to the
impurity-scattering $T$-matrix is shown by Fig.~2(c).
\begin{figure}[htbp]
  \begin{center}
  \psfig{file=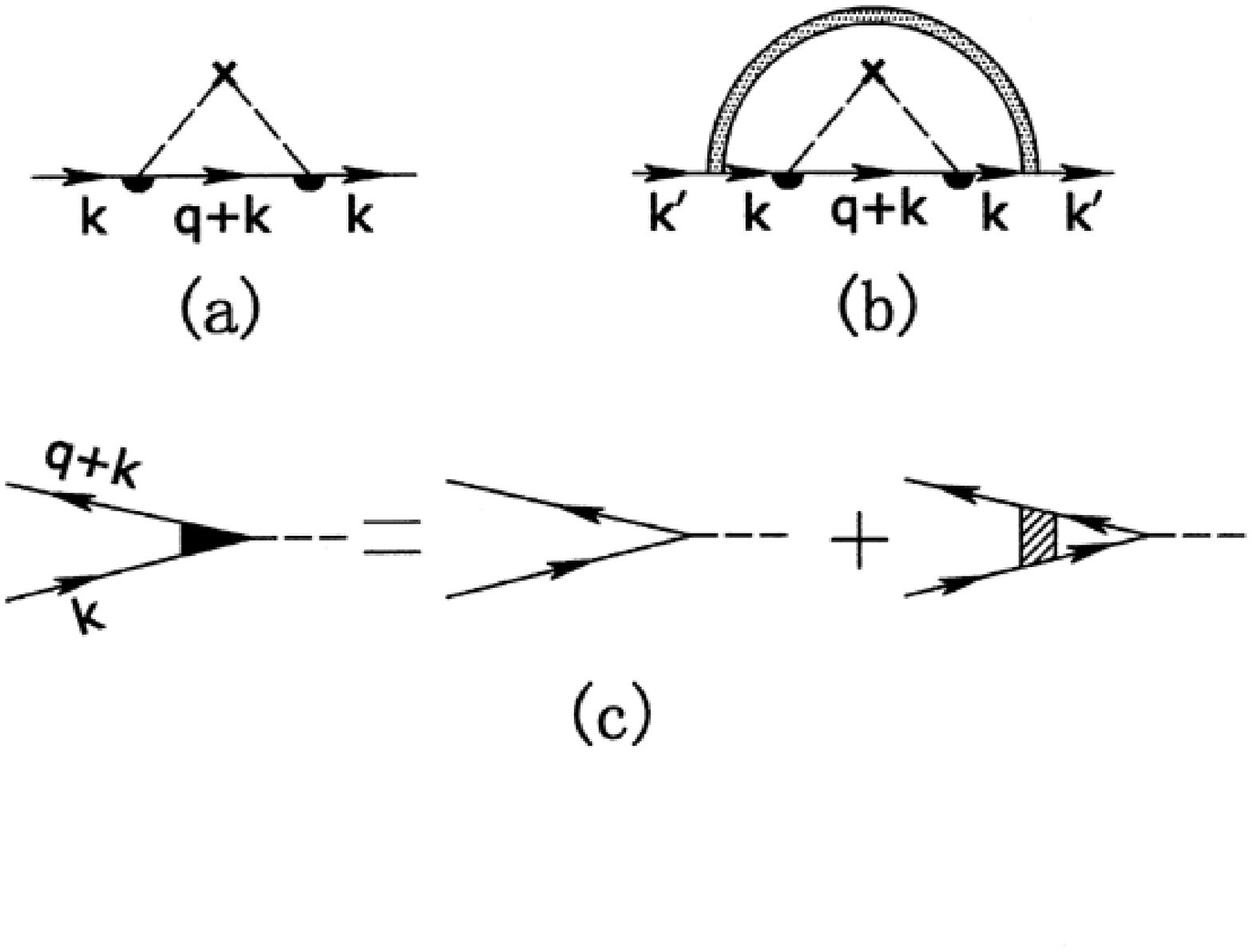,width=14cm,bbllx=12pt,bblly=291pt,bburx=570pt,bbury=621pt}
    \parbox{12.5cm}{\caption{\footnotesize The lowest-order self-energy diagrams with 0-mode diffuson
(a and b), and the vertex diagram (c). The shaded and grey blocks
represent, respectively, the 0-mode diffuson and non-singular
ladders. The dashed lines denote the impurity-scattering
$T$-matrix. The corresponding diagrams with $\pi$-mode diffuson
can be generated by the replacement of ${\bm q}\rightarrow{\bm
Q}+{\bm q}$ in these 0-mode diagrams.}}
     \end{center}
\end{figure}
Unlike the single-vertex-dressed diagram used in Ref.~\cite{23},
the present Figs.~2(a) and 2(b) include the additional diagrams
with {\it both} vertices dressed by the $\pi$-mode diffuson.
Similar vertex correction has been considered in the theory of
disordered interacting-electron systems~\cite{29}. The
vertex-corrected retarded $T$-matrix can be expressed by
\begin{equation}
{\bar T}^R({\bm q},\epsilon)_{\mu\mu'}=\sum_{\nu\nu'}{\cal J}({\bm
q},\epsilon)_{\mu\mu',\nu\nu'}^{RR}T^R(\epsilon)_{\nu\nu'},
\end{equation}
where $\mu,\mu',\nu$, and $\nu'$ are the indices of the Nambu
space, and the vertex function ${\cal J}({\bm q},\epsilon)^{RR}$
is given by
\begin{equation}
{\cal J}({\bm q},\epsilon)^{RR}={\cal I}+{\cal D}({\bm
q};\epsilon,\epsilon)^{RR}{\cal H}({\bm
q};\epsilon,\epsilon)^{RR}.
\end{equation}

In order to calculate the vertex function, we exploit the equation
for 0-mode diffuson in the RR-channel
\begin{equation}
{\cal D}({\bm q};\epsilon,\epsilon)^{RR}={\cal
W}(\epsilon,\epsilon)^{RR}+{\cal W}(\epsilon,\epsilon)^{RR}{\cal
H}({\bm q};\epsilon,\epsilon)^{RR}{\cal D}({\bm
q};\epsilon,\epsilon)^{RR},
\end{equation}
yielding
\begin{equation}
{\cal H}({\bm q};\epsilon,\epsilon)^{RR}={\cal
W}^{-1}(\epsilon,\epsilon)^{RR}-{\cal D}^{-1}({\bm
q};\epsilon,\epsilon)^{RR}.
\end{equation}
A substitution of Eq.~(4.12) into Eq.~(4.10) leads to
\begin{equation}
{\cal J}({\bm q},\epsilon)^{RR}={\cal D}({\bm
q};\epsilon,\epsilon)^{RR}{\cal W}^{-1}(\epsilon,\epsilon)^{RR}.
\end{equation}
Making use of Eqs.~(3.6) and (3.7), we get
\begin{eqnarray}
{\cal
W}^{-1}(\epsilon,\epsilon)^{RR}\approx\frac{\pi\rho_0}{2\gamma}\times
\left\{\begin{array}{cc}\tau_3\otimes\tau_3,~&{\rm for ~the ~Born ~limit},\\
     -\tau_0\otimes\tau_0,~&{\rm
     for ~the ~unitary ~limit}.\end{array}\right.
\end{eqnarray}
Substituting Eqs.~(3.29) and (4.14) into Eq.~(4.13), we obtain the
expression of the vertex function as (for $Dq^2\ll \gamma$ and
$|\epsilon|\ll \gamma$)
\begin{equation}
{\cal J}({\bm
q},\epsilon)^{RR}=\frac{2\gamma}{Dq^2-i2\epsilon}\Big(\tau_0\otimes
\tau_0-\tau_1\otimes \tau_1-\tau_2\otimes \tau_2-\tau_3\otimes
\tau_3\Big),
\end{equation}
which is suitable both near the Born and near the unitary limits.
Substituting Eqs.~(2.1) and (4.15) into Eq.~(4.9), we can easily
show that
\begin{equation}
{\bar T}^R({\bm q},\epsilon)=\sum_iJ({\bm
q},\epsilon)^{RR}_{ii}\tau_iT^R(\epsilon)\tau_i^*=0,
\end{equation}
indicating that both Figs.~2(a) and 2(b) have vanishing
contributions to the DOS near the Born or unitary limit.

\begin{center}
{\subsection*{\bf C. Contribution of the $\pi$-mode cooperon near
the UN limit}}
\end{center}

Near the UN limit, besides the diffusive 0-modes, the Goldstone
$\pi$-modes also contribute to the QI effect. The leading
self-energy diagrams with the $\pi$-mode cooperon can be obtained
from those in Fig.~1 by a replacement of ${\bm q}\rightarrow{\bm
q}+{\bm Q}$. Similarly with the case of 0-mode cooperon, the
contribution of $\pi$-mode cooperon to the DOS can be evaluated
via
\begin{equation}
\rho(\epsilon)_{\pi-{\rm coop}}=-\frac{2}{\pi}{\rm Im}\sum_{\bm
q}\sum_i\left[{\cal C}_\pi({\bm q};\epsilon,\epsilon)^{RR}{\cal
M}_\pi\right]_{ii},
\end{equation}
where
\begin{equation}
{\cal M}_\pi=\sum_{\bm k}G^R_{{\bm Q}-{\bm k}}\otimes(G^R_{\bm
k}KG^R_{\bm k}),
\end{equation}
with $K=l\tau_0/(2l-1)$ for the unitary limit. Expression (4.18)
can be readily evaluated by means of the global particle-hole
symmetry, yielding
\begin{equation}
{\cal M}_\pi=i\frac{l}{2\pi(2l-1)
v_fv_g\gamma}\Big(\tau_1\otimes\tau_1+\tau_3\otimes\tau_3\Big).
\end{equation}
Substituting Eqs.~(3.48) and (4.19) into Eq.~(4.17), we obtain
\begin{eqnarray}
\frac{\rho(\epsilon)_{\pi-{\rm
coop}}}{\rho_0}&=&\frac{2\pi}{\alpha (2l-1)}{\rm Re}\sum_{\bm
q}\frac{D}{Dq^2-i2\epsilon+2\delta}\nonumber\\&=&\frac{1}{2\alpha
(2l-1)}\ln\frac{\gamma}{\sqrt{\epsilon^2+\delta^2}}.
\end{eqnarray}
Comparing Eq.~(4.8) with Eq.~(4.20), one finds that the
contribution to DOS of the $\pi$-mode cooperon has an equal
magnitude but an opposite sign to that of the 0-mode cooperon in
the UN limit ($\delta\rightarrow 0$).

\vspace{0.5cm}
\begin{center}
{\subsection*{\bf D. Contribution of the $\pi$-mode diffuson near
the UN limt}}
\end{center}
\vspace{-0.8cm}

The leading self-energy diagrams containing the $\pi$-mode
diffuson are generated by replacing ${\bm q}$ by ${\bm Q}+{\bm q}$
in Figs.~2(a) and 2(b). Obviously, Eqs.~(4.9) and (4.13) are also
suitable for the $\pi$-mode diffuson. By exploiting Eq.~(3.48), we
can readily show that
\begin{equation}
{\cal J}_\pi({\bm
q},\epsilon)^{RR}=\frac{2\gamma}{Dq^2-i2\epsilon+2\delta}\Big(\tau_0\otimes
\tau_0+\tau_1\otimes \tau_1-\tau_2\otimes \tau_2+\tau_3\otimes
\tau_3\Big)
\end{equation}
and
\begin{eqnarray}
{\bar T}^R_\pi({\bm q},\epsilon)&=&\sum_iJ_\pi({\bm
q},\epsilon)^{RR}_{ii}\tau_iT^R(\epsilon)\tau_i^*\nonumber\\
&=&-i\frac{16\gamma}{\pi\rho_0}\frac{1}{Dq^2-
i2\epsilon+2\delta}\tau_0,
\end{eqnarray}
which are valid near the UN limit.

The contribution of the $\pi$-mode diffuson to the DOS is given by
\begin{equation}
\rho(\epsilon)_{\pi-{\rm diff}}=-\frac{n_i}{\pi}{\rm Im}\sum_{{\bm
k}{\bm q}}{\rm Tr}\left[{\bar T}^R_\pi({\bm q},\epsilon)G^R_{{\bm
Q}+{\bm k}}{\bar T}^R_\pi({\bm q},\epsilon)G^R_{\bm k}KG^R_{\bm
k}\right].
\end{equation}
Substituting Eqs.~(2.2) and (4.22) into Eq.~(4.23), and using
Eqs.~(2.6) and (A1)--(A4), one can easily show that
\begin{eqnarray}
\frac{\rho(\epsilon)_{\pi-{\rm
diff}}}{\rho_0}&=&\frac{32\pi\gamma}{\alpha(2l-1)}{\rm
Re}\sum_{\bm
q}\frac{D}{(Dq^2-i2\epsilon+2\delta)^2}\nonumber\\
&=&\frac{4}{\alpha(2l-1)}\frac{\gamma\delta}{\epsilon^2+\delta^2}.
\end{eqnarray}
Expression (4.24) is quite different from the logarithmic behavior
obtained in Ref.~\cite{23}, due to the additional contributions of
the self-energy diagrams with both the impurity-scattering
vertices dressed by the $\pi$-mode cooperon.

\begin{center}
{\subsection*{\bf E. Results of the QI correction to DOS}}
\end{center}

For generic Fermi surfaces, the QI correction to the DOS results
only from the contribution of 0-mode cooperon, i.e.,
\begin{equation}
\frac{\Delta\rho(\epsilon)}{\rho_0}=\frac{\rho(\epsilon)_{\rm
coop}}{\rho_0}=-\frac{1}{2\alpha\zeta}\ln\frac{\gamma}{|\epsilon|},
\end{equation}
where $\zeta$ is given by Eq.~(4.6). Therefore, the QI effect
gives rise to a logarithmic suppression to the quasiparticle DOS,
as predicted by Yashenkin et al.~\cite{23}. Equation (4.25) is
suitable near the Born or unitary limit. In the intermediate
region far from these two limits, however, the 0-mode cooperon may
contain some singular non-diagonal components, and thus a refined
theory is necessary for the generic situations. In addition, the
additional contributions of the self-energy diagrams with
non-singular ladders yield various renormalization factors in
these two limits.

Near the UN limit, the QI correction to the DOS is the sum of
$\rho(\epsilon)_{\rm coop}$, $\rho(\epsilon)_{\pi-{\rm coop}}$,
and $\rho(\epsilon)_{\pi-{\rm diff}}$, so that
\begin{equation}
\frac{\Delta\rho(\epsilon)_{\rm UN
}}{\rho_0}=\frac{1}{2\alpha(2l-1)}\left[-\ln\frac{\gamma}{|\epsilon|}
+\ln\frac{\gamma}{\sqrt{\epsilon^2+\delta^2}}
+\frac{8\gamma\delta}{\epsilon^2+\delta^2}\right].
\end{equation}
The above equation indicates that the quasiparticle DOS near the
UN limit can be subject to a positive correction due to the
contributions of the Goldstone $\pi$-modes. Upon approaching the
UN limit ($\delta$ is small enough), the contributions from the
0-mode and $\pi$-mode cooperons cancel out each other, so that the
DOS correction is given only by the contribution of the $\pi$-mode
diffuson, i.e.,
\begin{equation}
\lim_{\delta\rightarrow0}\frac{\Delta\rho(\epsilon)_{\rm UN
}}{\rho_0}=\frac{4\pi\gamma}{\alpha(2l-1)}\delta(\epsilon).
\end{equation}
This result can be used to account for the sharp peak found in the
numerical studies of the DOS~\cite{19,20}.

\vspace{0.5cm}
\begin{center}
{\section*{{\bf V. Summary}}}
\end{center}
\setcounter{section}{5} \setcounter{equation}{0} \vspace{-0.8cm}

We have extensively investigated the QI effect on the
quasiparticle DOS in a $d_{x^2-y^2}$-wave superconductor with
dilute nonmagnetic impurities. As in the study of disordered
normal metals~\cite{24}, the understanding of the diffusive modes
is essential for the investigation of the QI effect in a
superconductor. Through detailed derivations, we have obtained the
expressions for the diffusive modes both near the Born and near
the unitary limits. It is demonstrated that these Goldstone modes
may contain non-diagonal components in the intermediate region far
from these two limits. Therefore, a refined weak-localization
theory is necessary for the generic situations.

For generic Fermi surfaces, the QI effect results only from the
0-mode cooperon, yielding a logarithmic suppression to the
quasiparticle DOS near the Born or unitary limit. We show that
those non-trivial self-energy diagrams containing non-singular
ladders give rise to various renormalization factors of the DOS
correction for the Born and unitary limits.

Near the UN (unitary and nesting) limit, the QI effect comes not
only from the contribution of the 0-mode cooperon, but also from
those of the $\pi$-mode cooperon and diffuson. It is found that
the self-energy diagrams with both impurity-scattering vertices
corrected by the $\pi$-mode diffuson have an important
contribution to the DOS, which is proportional to
$\delta/(\epsilon^2+\delta^2)$. As a result, the quasiparticle DOS
is subject to a positive correction induced by the diffusive
$\pi$-modes.  Upon approaching the UN limit
($\delta\rightarrow0$), the contributions of the 0-mode and
$\pi$-mode cooperons cancel out each other (even including the
renormalization contributions induced by the non-singular
ladders), so that the DOS correction  becomes a $\delta$-function
of the energy. This result can be used to account for the resonant
peak found in the previous numerical studies~\cite{19,20}. A
similar sharp peak of electronic DOS has been found numerically in
a disordered 2D tight-binding model for the normal
state~\cite{30}. The appearance of this sharp peak can be also
explained by the contribution of Fig.~2(a) with the $\pi$-mode
diffuson (Fig.~2(b) has a vanishing contribution for the normal
state)~\cite{31}.

Like in the case of disordered interacting-electron
systems~\cite{29}, all the leading polarization diagrams
responsible for the QI effect on the quasiparticle conductitity
can be generated in the conserving approximation from the
lowest-order self-energy diagrams shown in Figs.~1 and 2. For the
normal state, it has been shown that the contributions of those
polarization diagrams with $\pi$-mode diffuson lead to an
antilocalization correction to the conductivity in the UN
limit~\cite{32}. How the QI processes related with the diffusive
$\pi$-modes affect the transport properties, such as electrical
and spin conductivities, in a $d$-wave superconductor is another
interesting problem, and will be studied in our future work.
\begin{center}
{\large {\bf Acknowledgment}}
\end{center}

This work was supported by the National Natural Science Foundation
of China under Grants No. 10274008 and No. 10174011, and the
Jiangsu-Province Natural Science Foundation of China under Grants
No. BK2002050 and BK2001002. DYX would like to acknowledge the
support of Grant No. G19980614 for State Key Programs for Basic
Research of China.
\renewcommand{\theequation}{A\arabic{equation}}
\vspace{0.5cm}
\begin{center}
{\large {\bf Appendix: Some useful mathematical formulas}}
 \end{center}
 \setcounter{equation}{0}

In this appendix, we present some mathematical formulas, which are
useful for the evaluations in the previous sections. By the
approach used in Ref.~\cite{33}, we can show that
\begin{equation}
\sum_{\bm k}\frac{\xi^2_{\bm k}\Delta^2_{\bm
k}}{(\gamma^2+\epsilon^2_{\bm k})^4}=\frac{1}{24\pi
v_fv_g\gamma^2},
\end{equation}
\begin{equation}
\sum_{\bm k}\frac{\xi^4_{\bm k}}{(\gamma^2+\epsilon^2_{\bm
k})^4}=\sum_{\bm k}\frac{\Delta^4_{\bm
k}}{(\gamma^2+\epsilon^2_{\bm k})^4}=\frac{1}{8\pi
v_fv_g\gamma^2},
\end{equation}
\begin{equation}
\sum_{\bm k}\frac{1}{(\gamma^2+\epsilon^2_{\bm
k})^n}=\frac{1}{(n-1)\pi v_fv_g\gamma^{2(n-1)}},~~{\rm
for}~n\geqslant2,
\end{equation}
and
\begin{equation}
\sum_{\bm k}\frac{\xi^2_{\bm k}}{(\gamma^2+\epsilon^2_{\bm
k})^n}=\sum_{\bm k}\frac{\Delta^2_{\bm
k}}{(\gamma^2+\epsilon^2_{\bm k})^n}=\frac{1}{2(n-1)(n-2)\pi
v_fv_g\gamma^{2(n-2)}},~~ {\rm for}~n\geqslant3.
\end{equation}
As an example, we shall prove Eq.~(A1). Using the Dirac-type
quasiparticle spectrum, and noting that there exist four gap
nodes, we have
\[
\sum_{\bm k}\frac{\xi^2_{\bm k}\Delta^2_{\bm
k}}{(\gamma^2+\epsilon^2_{\bm k})^4}=4\int\int
\frac{d\tilde{k}_fd\tilde{k}_g}{(2\pi)^2}\frac{v^2_fv^2_g\tilde{k}^2_f\tilde{k}^2_g}
{(\gamma^2+v^2_f\tilde{k}^2_f+v^2_g\tilde{k}^2_g)^4}.
\]
By means of the transformations of $p_f=\sqrt{v_f/v_g}\tilde{k}_f$
and $p_g=\sqrt{v_g/v_f}\tilde{k}_g$, the above equation can be
changed as
\begin{eqnarray}
\sum_{\bm k}\frac{\xi^2_{\bm k}\Delta^2_{\bm
k}}{(\gamma^2+\epsilon^2_{\bm k})^4}&=&4\int\int\frac{
dp_fdp_g}{(2\pi)^2}\frac{v^2_fv^2_gp^2_fp^2_g}{[\gamma^2+v_fv_g(p^2_f+p^2_g)]^4}\nonumber\\
&=&\frac{1}{2\pi^2v_fv_g\gamma^2}\int^{x_0}_{0}dx\frac{x^2}{(1+x)^4}\int^{2\pi}_{0}d\theta
\cos^2\theta\sin^2\theta,\nonumber
\end{eqnarray}
where $x_0 =v_fv_gp^2_0/\gamma^2$ with $p_0\sim1/a$. For the
weak-disorder case ($\gamma$ is small enough) under considered, we
can set $x_0=\infty$. Thus, a completion of the integrals over $x$
and $\theta$ in the above equation immediately yields Eq.~(A1).

Another useful formula is given by
\begin{equation}
\frac{1}{2}\sum_iC_{ii}{\rm
Tr}(\tau_iA\tau_iB)=\sum_i({\cal CM})_{ii},
\end{equation}
where ${\cal C}=\sum_iC_{ii}\tau_i\otimes\tau_i$, and ${\cal
M}=A\otimes B$ with $A$ and $B$ being any linear superimpositions
of $\tau_i$ ($i$=0,1,2,3). In order to prove Eq.~(A5), we assum
that $\tau_iA=\sum_jx_{ij}\tau_j$ and
$\tau_iB=\sum_ky_{ik}\tau_k$. Then we have
\begin{equation}
\frac{1}{2}\sum_iC_{ii}{\rm
Tr}(\tau_iA\tau_iB)=\frac{1}{2}\sum_{ijk}C_{ii}x_{ij}y_{ik}{\rm
Tr}(\tau_j\tau_k)=\sum_{ijk}C_{ii}x_{ij}y_{ik}\delta_{jk}=\sum_{ij}C_{ii}x_{ij}y_{ij}
\end{equation}
and
\begin{equation} {\cal CM}=\sum_i
C_{ii}(\tau_iA)\otimes(\tau_iB)=\sum_{ijk}C_{ii}x_{ij}y_{ik}\tau_j\otimes\tau_k.
\end{equation}
From Eq.~(A7), it follows that
\begin{equation}
\sum_i({\cal CM})_{ii}=\sum_{ij}C_{ii}x_{ij}y_{ij}.
\end{equation}
A combination of Eq.~(A6) with Eq.~(A8) immediately leads to
Eq.~(A5).

\end{document}